# How the medium shapes the message: Printing and the rise of the arts and sciences

[1], C. Jara-Figueroa[a], Amy Z. Yu[a], and César A. Hidalgo[a,b]

[a]The MIT Media Lab, Massachusetts Institute of Technology
[b]to whom correspondence should be addressed: hidalgo@mit.edu

**Abstract**

Communication technologies, from printing to social media, affect our historical records by changing the way ideas are spread and recorded. Yet, finding statistical instruments to address the endogeneity of this relationship has been problematic. Here we use a city's distance to Mainz as an instrument for the introduction of the printing press in European cities, together with data on nearly 50 thousand biographies, to show that cities that adopted printing earlier were more likely to be the birthplace of a famous scientist or artist in the years after the introduction of printing. At the global scale, we find that the introduction of printing is associated with a significant and discontinuous increase in the number of biographies available from people born after the introduction of printing. We bring these findings to more recent communication technologies by showing that the number of radios and televisions in a country correlates with the number of performing artists and sports players from that country that reached global fame, even after controlling for GDP, population, and including country and year fixed effects. These findings support the hypothesis that the introduction of communication technologies shift historical records in the direction of the content that is best suited for each technology.

Communication technologies have been argued to affect society by shaping institutions [1–5], changing the way people think [6–8], biasing the content that people transmit and record [9–12], and facilitating access to knowledge [13–18]. Yet, statistically testing the effects of communication technologies in society has been challenging because of both, a lack of structured data and a lack of statistical instruments that we can use to address the endogeneity of the relationship between communication technology and content. Do new technologies promote the creation of specific types of content? Do external factors, such as cultural revolutions, change both the content and the technologies needed to diffuse that content? Or does the emergence of new content create the demand needed to make the technologies that facilitate its diffusion?

Here we use data on nearly forty thousand biographies to study how three historical changes in media—printing (1450-1550), radio (1890-1950), and television (1950-1970)—abruptly changed the occupations of the people associated to the biographies we find in modern historical records. What these changes in media have in common is that they had global—or at least continental—impact, and that they had relatively short adoption times [11–14, 18].

First, we use changepoint analysis to identify discontinuities in the time series of the per-capita number of biographies and find that the only discontinuity identified by this statistical method, between 500 and 1650, coincides with the introduction of Gutenberg's movable type printing press. Then, we use distance to Mainz as an instrumental variable to measure the effect of the adoption of printing in a city's ability to produce memorable scientists and artists. While the ability of a city to produce famous scientist and artists may depend on a large number of variables, such as its population and wealth, that city's distance to Mainz can only affect the number of famous people born in that city through the adoption of printing. Hence, a city's distance to Mainz is an exogenous form of variation that increases its likelihood of being an early adopter of printing technology. In fact, between 1450 and 1500, printing diffused in concentric circles as printers set out to establish presses in other cities [17]. Therefore, it is not surprising that distance to Mainz has been used to establish the impact of the printing press in economic growth [17], and on the spread of protestantism [19]. Our two stage least square estimate using distance to Mainz as an instrument confirms the hypothesis that cities that adopted printing earlier produced famous scientists and famous artists earlier, indicating that printing promoted the emergence of a scientific and artistic elite.

To bring these findings to other eras we explore the occupations of the famous biographies associated with each communication technology (writing, printing, radio, and television), finding that each era is characterized by its own composition of famous occupations. Finally, we use data at the country level on the number of radios and the number of televisions available each year in each country [20] to show that countries with more radios and with more televisions, were the birthplace of more famous artists and sports players, even after controlling for GDP, population, and including country and year fixed effects. The same is not true for occupations such as scientists, that experienced an increase with the rise of printing, but not with radio or television. These results support the idea that communication technologies shape our historical records by biasing their content towards the type of information best suited for the media of each time.

## Data and methods

We use biographical data from two sources: the *Pantheon 2.0* dataset and the *Human Accomplishments* dataset [21]. The *Human Accomplishments* dataset [21] contains more than three thousand biographies of individuals from the Arts and Sciences that are recorded in authoritative printed texts in six different languages. The Pantheon 2.0 is an extension to the Pantheon 1.0 dataset [22], which is a peer-reviewed dataset containing the 11,337 biographies that had a presence in more than 25 different language editions of Wikipedia as of May 2013 (for more details see [22]). Pantheon 2.0 extends the 1.0 version to include people with more than 15 languages in Wikipedia as of July 2016. The Pantheon 2.0 dataset associates each biography with a city of birth, a date of birth, and an occupation. The place and date of birth are taken from the information provided in each biography's Wikipedia infobox, complemented with the information present in each biography's Wikidata page. The city of birth is obtained by associating each place of birth to a city from the GeoNames database [23] (we use cities over 5000 people). The occupation is the result of a classifier that extracts features from the biography's text on Wikipedia, and was trained on the Pantheon 1.0 dataset (for more information see SI). For our analysis, we aggregate occupations into 8 categories: political leaders, religious figures, scientists, performing artists, fine artist, humanities, and sports. More details on the data and the occupations can be found in SI and *pantheon.media.mit.edu*.

Both of these datasets belong to a recent stream of literature focused on the use and development of quantitative methods to explore historical patterns [21, 22, 24–27]. Examples of these studies include the evolution of language and ideas as recorded in printed books [25], patterns of historical migration [28], the importance of language translations in the global diffusion of information [29], the emotional content of global languages [30], and the dynamics of fame [22, 24, 31]. These studies are made possible thanks to new digital sources such as Wikipedia, Freebase, Wikidata, and digitized books [32].

Data on technology adoption comes from two sources. For printing, we use the *Incunabula Short Title Catalogue* [33], a dataset comprising all books printed between 1450 and 1500. This data is suitable to understand the adoption of printing because the period 1450–1500 (the "first infancy" of printing [34]) was a period of rapid expansion of printing among European cities, with the price of books falling by nearly two thirds [35, 36]. For radio and for television we use the Historical Cross-Country Technology Adoption (HCCTA) dataset [20], a dataset collected to analyze the adoption patterns of some of the major technologies introduced in the past 250 years. The HCCTA dataset also has historical GDP and population information.

Population data comes from two sources. At the global level we use data from the historical world population estimates of the US Census Bureau [37], which reports an aggregated dataset of world population estimates starting from the year 10,000 BC. At the city level we use a dataset on population of urban settlements from 3700 BC to AD 2000 [38]. We matched the cities present in [38] with the ones in [23] based on their name and coordinates.

All of our data sources include limitations that need to be considered when interpreting our results. Therefore, we emphasize the need to interpret our results as valid only in the narrow context of the sources used to compile these datasets. The Pantheon 2.0 dataset has all the biases inherent in using Wikipedia as a primary data source [22], therefore our results should be interpreted as a reflection of the historical records that are representative only of the people who edit the more than 250 language editions of Wikipedia—a literate and digitally empowered elite of knowledge enthusiasts [39]. The Human Accomplishments dataset is compiled from printed encyclopedias in six different languages, and is representative of the people who participated in the creation of printed encyclopedias. An extended discussion of the biases and validation of this dataset can be found in Human Accomplishments [21].

## Results

We study the global effects of printing by looking at the number of biographies $B$ born in a given time window $\Delta T$ for both the Pantheon 2.0 and the Human Accomplishment datasets. To make these estimates comparable across time we normalize $B$ by the average world population $N$ and its respective time window, defining $m = \frac{B}{N\Delta T}$. $m$ is an estimate of the per-capita number of births occurring in a given year that resulted in a biography that is prominently recorded in each of our datasets. $m$ is measured in units of births per year per billion people in the world, or [bpyb].

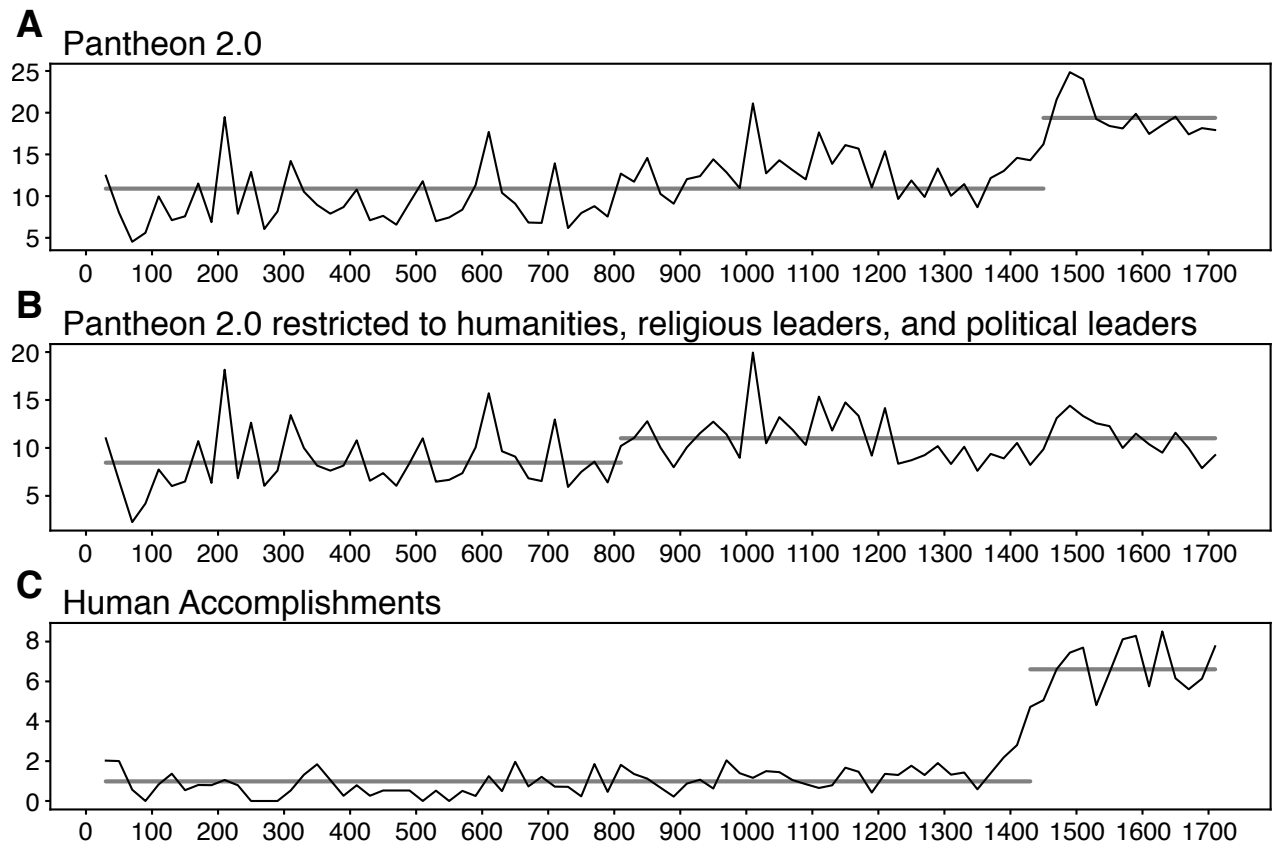

**Fig. 1.** Per-capita births of globally memorable people ($m$) measured using a 20-years time window for **A** the full Pantheon 2.0 dataset, **C** Pantheon 2.0 dataset restricted to **government**, **humanities**, and **religion**, and **C** the Human Accomplishments dataset. The horizontal lines correspond to the average for each part of the time series, as detected by the changepoint analysis.

Figures 1 A and C show the temporal evolution of $m$ as observed in both datasets between 1 and 1730, using a 20-years time window. In both datasets $m$ is constant for the 1,500 years preceding the introduction of the movable type press but increases by more than 50% together with the introduction of printing according to changepoint analysis [40]. Furthermore, we find that once we limit the Pantheon 2.0 dataset to only political and religious leaders–the two categories that were prominently present in the dataset prior to printing (see Figure 2 A)–there is no statistically significant break (see Figure 1-B). This suggests that the increase in the number of per-capita biographies observed with the rise of printing was the result of a new cultural elite that included artists and scientists and that was not prominent prior to printing technology.

Next, we address endogeneity concerns by using the distance to Mainz as an instrumental variable [17, 19]. We use two different variables, calculated at the city level, to quantify early adoption of printing: a dummy variable for adopting printing between 1450 and 1500 ($D_{\text{printer}}$), and the year of the first printed book (for a third empirical specification see SI). Our first instrumental variable analysis uses a city's distance to Mainz as an instrument for adopting printing between 1450 and 1500, obtained from the Incunabula Short Title Catalogue [33]. As dependent variables we use the number of scientists, artists, and political leaders born in each city between 1450 and 1550, right after the invention of printing (for the other occupations see SI). If the introduction of printing gave rise to a cultural elite of scientists and artists, we should observe a significant effect in the number of scientists and artists, but not in the number of political leaders, who were already prominent prior to printing. Table 1 shows the first stage, and the OLS and IV estimators for each dependent variable. We see that adopting printing earlier, $D_{\text{printer}}$, has a significant, positive, and strong effect on the number of scientists and artists, but does not have a significant effect on the number of political leaders, meaning that early adopters of printing were the birthplace of more scientists and artists than late adopters. For our second instrumental variable analysis, we use distance to Mainz as an instrument for the year of the first printed book. This analysis is restricted to cities that adopted printing, so the number of observations decreases considerably. As dependent variables we use the year of the first recorded scientist born in each city, the year of the first recorded artist born in each city, and the year of the first political leader born in each city (for the other occupations see SI). Table 2 shows that the year of the first printed book has a significant, positive, and strong effect on the production of scientists and artists, but does not have a significant effect on the production of political leaders, meaning that cities that adopted printing earlier were the birthplace of famous scientists and artists earlier than late adopters.

Next, we look at the occupations associated with the biographies in Pantheon 2.0 dataset at each point in time. Figure 2 shows the fraction of biographies corresponding to each occupation category for each period punctuated by the technological breaks we focus on in this work. First, comparing figures 2 A and B we see that printing is associated with an increase in the fraction of painters, composers, and scientists (such as physicists, mathematicians, and astronomers), and a decrease in the fraction of religious figures. Second, radio, a technology at about the same time as film, (Figures 2 B and C) was accompanied by a shift towards the performing arts and an increase in the number of actors, singers, and musicians. Finally, with the introduction of television (Figures 2 C and D) we see the rise of sports players—such as soccer players, basketball players, and race-car drivers—and the consolidation of performing artists.

Table 3 shows the results of regressing the number of performing artists born each year in each country, the number of sports players, and the number of scientists, on the number of radios $n_{rads}$, and the number of televisions $n_{tvs}$. Since we are dealing with panel data, all models in Table 3 include country fixed effects and year fixed effects. The results show that, after controlling for GDP and population, countries that had more radios and television were the birthplace of more famous performing artists, and sports players. Moreover, radios and televisions have no correlation with the number of scientists born each year in each country, in fact the difference in the explanatory between models (8) and (9) from Table 3 is not significant (p-value$\sim$ 0.19), while the difference between models (2) and (3), and (5) and (6) are significant (p-value$\sim 10^{-16}$).

| | *Dependent variable:* | | | | | | | |
|---|---|---|---|---|---|---|---|---|
| | $D_{\text{printer}}$ | | number of scientists | | number of artists | | number of political leaders | |
| | *probit* | *First Stage* | *OLS* | *IV* | *OLS* | *IV* | *OLS* | *IV* |
| | (1) | (2) | (3) | (4) | (5) | (6) | (7) | (8) |
| $\log(d_{\text{Mainz}})$ | $-0.273^{***}$ | $-0.022^{***}$ | | | | | | |
| | (0.033) | (0.003) | | | | | | |
| $D_{\text{printer}}$ | | | $0.089^{***}$ | $0.222^{***}$ | $0.202^{***}$ | $0.302^{***}$ | $0.188^{***}$ | 0.130 |
| | | | (0.004) | (0.043) | (0.007) | (0.067) | (0.009) | (0.081) |
| constant | $-0.014$ | $0.186^{***}$ | $0.002^{**}$ | $-0.003$ | $0.003^{**}$ | $-0.0002$ | $0.010^{***}$ | $0.012^{***}$ |
| | (0.212) | (0.018) | (0.001) | (0.002) | (0.001) | (0.003) | (0.002) | (0.003) |
| Observations | 5,335 | 5,335 | 5,335 | 5,335 | 5,335 | 5,335 | 5,335 | 5,335 |
| Log Likelihood | $-804.19$ | | | | | | | |
| AIC | 1,612 | | | | | | | |
| Residual SE (df = 5333) | | 0.187 | 0.061 | 0.066 | 0.100 | 0.102 | 0.124 | 0.124 |
| $R^2$ | | 0.012 | 0.070 | | 0.125 | | 0.075 | |
| Adjusted $R^2$ | | 0.012 | 0.070 | | 0.125 | | 0.075 | |
| F Statistic (df = 1; 5333) | | $66.57^{***}$ | $399.43^{***}$ | | $762.57^{***}$ | | $435.08^{***}$ | |
| *Note:* | | | | | | $^{*}$p<0.1; $^{**}$p<0.05; $^{***}$p<0.01 | | |

**Table 1.** Instrumental variable analysis of the effect of printing on European cities. Here, we use the distance to Mainz ($d_{\text{Mainz}}$) as an instrument for a dummy variable for adopting printing between 1450 and 1500, $D_{\text{printer}}$. We use a two stage least squared regression to estimate the effect of adopting printing between 1450 and 1500 on the number of scientists, artists, and political leaders born in each city between 1400 and 1550. All dependent variables are in logarithmic scale.

| | *Dependent variable:* | | | | | | |
|---|---|---|---|---|---|---|---|
| | year of first printer | year of first scientist | | year of first artist | | year of first political leader | |
| | *First Stage* | *OLS* | *IV* | *OLS* | *IV* | *OLS* | *IV* |
| | (1) | (2) | (3) | (4) | (5) | (6) | (7) |
| $d_{\text{Mainz}}$ | $0.010^{***}$ | | | | | | |
| | (0.003) | | | | | | |
| year of first printer | | $4.870^{***}$ | $12.900^{**}$ | $6.552^{***}$ | $12.956^{**}$ | $3.036^{*}$ | $-1.955$ |
| | | (1.425) | (5.655) | (1.740) | (6.194) | (1.809) | (4.988) |
| $\log(pop)$ | $-3.280^{**}$ | $-18.507$ | $-5.089$ | $-37.476^{*}$ | $-38.083$ | $-64.311^{**}$ | $-68.651^{**}$ |
| | (1.644) | (19.035) | (24.651) | (21.950) | (24.072) | (26.015) | (27.613) |
| constant | $55.853^{***}$ | 294.220 | $-69.230$ | 383.895 | 215.241 | $719.812^{**}$ | $908.366^{***}$ |
| | (16.207) | (203.821) | (346.180) | (231.540) | (297.534) | (275.779) | (337.974) |
| Observations | 82 | 73 | 73 | 70 | 70 | 77 | 77 |
| $R^2$ | 0.132 | 0.163 | | 0.202 | | 0.113 | |
| Adjusted $R^2$ | 0.111 | 0.139 | | 0.178 | | 0.089 | |
| F Statistic | $6.032^{***}$ | $6.833^{***}$ | | $8.496^{***}$ | | $4.732^{**}$ | |
| | (df = 2; 79) | (df = 2; 70) | | (df = 2; 67) | | (df = 2; 74) | |
| *Note:* | | | | | $^{*}$p<0.1; $^{**}$p<0.05; $^{***}$p<0.01 | | |

**Table 2.** Instrumental variable analysis of the effect of printing on European cities that adopted printing. Here, we use the distance to Mainz ($d_{\text{Mainz}}$) as an instrument for the year of the first printing press in each city, and use it to estimate the effect of printing in the birth of scientists, artists, and political leaders.

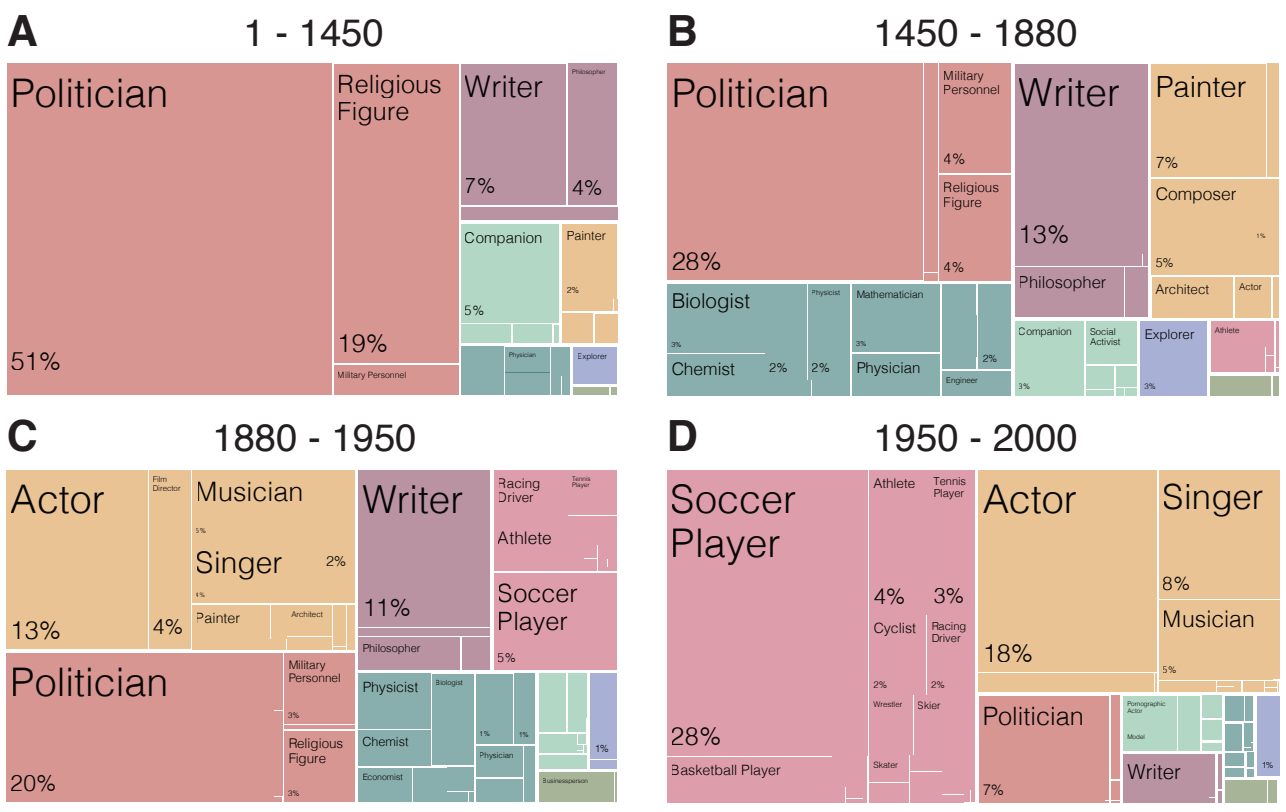

**Fig. 2.** Composition of occupations for the biographies in Pantheon 2.0 for the period between **A** 1-1450, **B** 1450-1880, **C** 1880-1950, and **D** 1950-2000.

# Discussion

In this paper we studied how communication technologies shape our historical records by biasing the types of information a society can transmit and record. As noted earlier, we are by no means the first ones to debate this question [6–18, 27]. Nevertheless, here we contribute to that debate by using two large biographical datasets, together with an instrumental variable, to quantitatively study some of the hypothesized consequences of changes in media.

The first technology studied in this work was Gutenberg's movable type printing press. Printing spread quickly across Europe starting with its introduction in Mainz, Germany close to 1450. In less than forty years the price of books fell by two-thirds, transforming the ways in which ideas and data were disseminated, and thus the conditions for intellectual work. The social and economic effects of printing were discussed in detail by the printing historian Elizabeth Eisenstein [13, 14] who argued that printing not only changed the number of books printed during the Renaissance, but also who the authors of these books were and what these books were about. According to Eisenstein, printing shifted the production of books away from religious texts—which were a staple output of the scribal culture that preceded printing—and towards art and science books. Printing enabled the proliferation of resources for artists and scientists, which, together with the new media that enabled scientists and artists to best disseminate their work, made the cities what were early adopters of printing better at producing scientists and artists, a fact that we observe in our analysis.

| | *Dependent variable:* | | | | | | | | |
|---|---|---|---|---|---|---|---|---|---|
| | number of performing artists | | | number of sports players | | | number of scientists | | |
| | (1) | (2) | (3) | (4) | (5) | (6) | (7) | (8) | (9) |
| $\log(n_{rads})$ | 0.037*** | | 0.031*** | 0.027*** | | 0.029*** | 0.003 | | −0.003 |
| | (0.005) | | (0.005) | (0.004) | | (0.004) | (0.004) | | (0.004) |
| $\log(n_{tvs})$ | 0.079*** | | 0.072*** | 0.040*** | | 0.042*** | −0.001 | | −0.007 |
| | (0.009) | | (0.009) | (0.006) | | (0.006) | (0.006) | | (0.006) |
| $\log(gdp)$ | | 0.477*** | 0.312*** | | 0.091* | −0.033 | | 0.218*** | 0.235*** |
| | | (0.069) | (0.067) | | (0.050) | (0.050) | | (0.043) | (0.044) |
| $\log(pop)$ | | −0.283*** | −0.158** | | −0.135** | −0.041 | | −0.044 | −0.056 |
| | | (0.070) | (0.068) | | (0.053) | (0.052) | | (0.044) | (0.044) |
| constant | −0.144 | 0.154 | −0.310 | −0.198*** | 0.965** | 0.599 | −0.017 | −1.046*** | −1.001*** |
| | (0.099) | (0.573) | (0.538) | (0.052) | (0.377) | (0.366) | (0.101) | (0.362) | (0.364) |
| fixed effects: | | | | | | | | | |
| country | ✓ | ✓ | ✓ | ✓ | ✓ | ✓ | ✓ | ✓ | ✓ |
| year | ✓ | ✓ | ✓ | ✓ | ✓ | ✓ | ✓ | ✓ | ✓ |
| Observations | 3,011 | 3,011 | 3,011 | 3,011 | 3,011 | 3,011 | 3,011 | 3,011 | 3,011 |
| $R^2$ | 0.638 | 0.622 | 0.645 | 0.576 | 0.551 | 0.578 | 0.539 | 0.556 | 0.557 |
| Adjusted $R^2$ | 0.615 | 0.599 | 0.623 | 0.549 | 0.523 | 0.552 | 0.511 | 0.529 | 0.529 |
| F Statistic | 28.49*** | 26.65*** | 29.10*** | 21.97*** | 19.90*** | 21.92*** | 18.98*** | 20.29*** | 20.09*** |
| *Note:* | | | | | | | | | *p<0.1; **p<0.05; ***p<0.01 |

**Table 3.** We explore the correlation between the number of radios and televisions a country has each year, with the number of performing artists, sports players, and scientists born each year, after controlling for GDP, population, and adding country fixed effects and year fixed effects. Reported errors correspond to robust standard errors. All quantities are calculated yearly between 1850 and 1973.

Not every historian, however, supports the importance of the introduction of the movable type press. An often cited argument is that the economic effects of the printing press were limited to the automation of the printing of books, a very small economic activity [41]. The effects of printing, however, should be understood not by looking at its effect on automating a single industry, but by looking at its effects on the spread of ideas.

Technological change is driven by sharing and recombining ideas [16, 42, 43]. Therefore, innovation depends on the cost of accessing existing knowledge [15]. Mokyr notes that the printing press was one of the most important “access cost-reducing inventions” in history. By changing the way ideas were spread, printing boosted the sciences and arts by facilitating a “combinatory intellectual activity” [13–15, 44]. Furthermore, other historians have argued that the introduction of the printing press in European cities promoted the accumulation of human capital, and played a key role in the evolution of new business practices [13–15, 44, 45], leading to higher growth rates [17]. In economic terms, cities that adopted printed earlier benefited from localized spillovers in human capital accumulation, leading to a higher chance of producing a successful scientist or artist. Our results show precisely that, since once we limit the dataset to only political and religious leaders, we see no significant break in the per-capita births of globally memorable people (see Figure 1). Moreover, we find that printing had no effect in a city’s ability to produce a successful political leader (see Tables 1 and 2).

The second and third changes in media studied in this work were the introduction of radio and film in the early twentieth century, and the introduction of television in the second half of the twentieth century. Both of these changes were also quick. The introductions of radio and television have been discussed by scholars such as Marshall McLuhan, who used the radio as an example of what he called a “hot” media [12]—media that deeply engage a single sense—and Neil Postman, who discussed the societal effect of television in the United States by arguing that television was a type of media that strongly favored entertainment content [11]. Postman discusses how television not only created new types of content, such as sports, but also affected existing types of content. For example, television changed the format in which political debates were carried out, and thus changed political discourse [11]. Following to Postman’s argument, because television favored entertainment it should be associated with the biographies of performers, such as actors and singers, and with biographies of sports players, but not with biographies of scientists. Our results show that even after controlling for population and gdp, the number of televisions and radios a country has, correlates with the number of performers and sports players born in each country. Moreover, the correlation between television and sports players is stronger than between radios and sports players (see Table 3), something that points at how interdependent television and sports have been from the beginning. In fact, the 1936 Summer Olympics, the first Olympic game to be broadcast on television, happened less than ten years after the first television station began its transmissions, and a year before the first televised theater play. In some sense, television "invented" sports, since due to its immediate nature [11] it favored entertainment content that also had an immediate nature.

Here, we have focused on three major changes in media. Something these changes in media have in common is that the changes in the content that was recorded following each one, always favored the occupations of those who can best express their work using the new media. The rise of printing, for instance, was followed by more composers recorded in our historical records, but not by more singers or musicians (see Figure 2). In the same way, film promoted the actor over the play-writer. In fact, changes in media helped include in our historical records occupations that had long existed, but that were not prominently recorded. For example, both Ancient Greek and Elizabethan English playwrights employed actors, but performers were not recorded in the absence of a technology capable of recording performances—such as film.

Finally, we note that observations for most recent years need to be interpreted carefully for two reasons. First, the more recent biographies in our dataset contain a mix of characters that are memorable (e.g. Barack Obama, as the first African American president of the United States) with characters whose presence in today's biographical records may not necessarily be long lasting (like teen pop icons and reality show celebrities). So the picture obtained for recent decades is not the one we expect to be representative of those decades in the future. Nevertheless, we can safely assume that this issue does not affect our historical data prior to the twentieth century, since these transient effects should not last for centuries after a person's death. Second, we note that data for the most recent years is also affected by differences in the life cycle of an individual's memorability, since individuals with different careers peak at different ages. Soccer players, for example, peak around their late twenties or early thirties [46], so our dataset should contain all soccer players born in the 1950s that became memorable as players. Political leaders and scientists, on the other hand, often become globally memorable much later in life [47], and hence, we may be missing some influential individuals who are yet to reach global recognition. Both of these effects imply that fifty years from now the fraction of our biographical records allocated to sports players will be smaller than what we observe in our data today. In other words, we expect the focus of history to adjust as time continues to elapse.

Prior to printing, history was limited to the most powerful institutional elites in the world. Now, we live in a world in which history is almost personalized, since billions of individuals now leave traces that could be used to reconstruct biographical data through personal acts of communication (emails, text messages, and social media posts). Of course, this does not mean that everyone will become memorable, but maybe that memorability will now have a chance to spread over a wider number of people who may now enjoy intermediate levels of memorability and fame. This is an effect that has already been observed in the context of creative industries [48]. Going forward, the rise of digitized historical records will help us continue the statistical study these and other hypotheses.

## Acknowledgments

All authors acknowledge support from the MIT Media Lab Consortia and from the Metaknowledge Network at the University of Chicago. We thank comments from Manuel Aristaran, Albert-Laszlo Barabasi, Jeffrey Ravel, Domink Hartmann, Jermain Kaminski, William Powers, Iyad Rahwan, and Ethan Zuckerman.

# How the medium shapes the message: Printing and the rise of the arts and sciences Supplementary Material

[1], C. Jara-Figueroa[a], Amy Z. Yu[a], and César A. Hidalgo[a,b]

[a]The MIT Media Lab, Massachusetts Institute of Technology
[b]to whom correspondence should be addressed: hidalgo@mit.edu

## Summary of data sources

### Pantheon

The Pantheon 2.0 dataset contains all the biographies that had a presence in more than 15 language editions of Wikipedia as of July 2016. As a starting point, we collected all the articles that belong to the WikiProject Biography from the English Wikipedia, and complemented this with all the *instances of* of the class *human* from Wikidata. Next, we use the Wikipedia API to collect the number of language editions for all biographies, and select only those biographies present in more than 14 language editions.

The articles within the scope of the WikiProject Biography, however, are not guaranteed to be about individual persons. For example, articles about music bands or duos, lists of monarchs, or terrorist groups are sometimes within the scope of the WikiProject Biography. To filter out the articles that do not correspond to a biography of a single person we manually check all the articles for which: i) we cannot find a *gender*, ii) the title that starts with the word "list," or that includes either "and" or "&", iii) contain either the word "band," "duo," or "group" after the verb "to be" in the first sentence of the description, and iv) the verb "to be" from the first sentence appears in its plural from.

Next, we assigned each biography to their year of birth and their place of birth using the Wikipedia Infobox present in most of the biographies. When the infobox was no available we used Wikidata. The granularity of the place of birth varies across briographies, with some biographies been assigned to countries, others to cities, and others to precise locations with cities (for example, a palace within a city). To use a controlled set for the cities, we associate each place of birth to a city from the GeoNames database [1]. In particular, we use *cities over 20000 people* (defined in GeoNames as settlements with a present day population of more than 20000 people). The assignment is done by matching the name of the place of birth with the name of the city, within a 30km radius. When this method failed, the place of birth was manually assigned. Countries, regions, and continents were not assigned to any city.

Finally, we describe the method used to obtain the occupations associated to the biographies in Pantheon 2.0. The occupations associated with characters in Pantheon are meant to capture the way the character is recorded in our historical records. Many characters hold an occupation that is not what they are remembered for. For example, Margaret Thatcher is a chemist and a lawyer, but she is recorded in our collective memory as the longest-serving British Prime Minister of the 20th century, and the only woman to have held that office so far; she is recorded as a politician. The available databases that contain information about historical characters, such as Freebase, Wikidata, and DBpedia, fail to associate a character to a single, most relevant, occupation.

Pantheon 2.0 follows a similar hierarchical occupation classification that Pantheon 1.0—e.g. Physics and Biology are branches of Natural Science, just like Natural Science is a branch of Science. Pantheon 1.0 has 88 different occupations at the lowest level of aggregation. Pantheon 2.0 adds 13 new occupations: youtuber, including people such as PewDiePie who are YouTube celebrities, occultist, including people such as Nostradamus who are said to have paranormal powers, inspiration, including people such as Vlad the Impaler who are famous for serving as inspiration for fictional characters, and the 10 new sports categories, badminton player, rugby player, handball player, bullfighter, volleyball player, pc gamer, poker player, go player, fencer, and table tennis player. A full list of all the occupations can be found in Table 6.

We use a SVM classifier, trained on the Pantheon 1.0 dataset, to classify the biographies of Pantheon 2.0. Since 13 new occupations were added in Pantheon 2.0, we complement Pantheon 1.0 with manually classified characters for each new occupation in order to build the training set.

The features we select for the classifier are drawn from Wikidata and Wikipedia, and are meant to characterize the character's main field of contribution. We select the following features:

- Infobox type: The type of the infobox templates used in the wikipage associated with the character. We filter out all the infoboxes that do not correspond to a biography.
- Wikidata occupations: For each character we collect all the values from the property "occupation" from Wikidata. We manually created a controlled vocabulary of 350 occupations—e.g. design engineer is mapped to engineer, marine biologist to biologist, etc.
- Extract words: We get the top 5 most frequent words from the character's extract, selected from a manually curated list of 750 words that are meant to capture the character's work. For example: backstroke, summit, reign, contribution, youtube, football, testament, etc.

When testing the classifier by setting aside 10% of the data, we get $\sim$ 93% success rate. The success rate is slightly higher in the second level of aggregation ($\sim$ 95%), which is the level we use to obtain the results reported in this work. Pantheon 2.0 classifies characters into 8 *domains*, 27 *industries*, and 101 *occupations*. For our analysis, we aggregate occupations into categories distinguishing mainly between politicians and religious figures, and between arts and performing arts. The *arts* domain is split into two groups *performing arts* (including *dance*, and *film and theater* industries, plus all occupations from the *music* industry except for *composer*) and *arts* (including *design* and *fine arts* industries and the *composer* occupation). The *religion* industry is grouped by itself, and all the other industries under the *institutions* domain are grouped together under *government*. The *team sports* industry is considered under *sports*. The *science and technology*, and *humanities* industries remain unchanged. Finally, *individual sports*, along with the domains *business and law*, *exploration*, and *public figure* are grouped together as *other*. Table 6 shows a summary of the aforementioned aggregation. We must note that we are not loosing meaningful information by creating the category *other*. The three largest occupations aggregated under *other*—*tennis player*, *social activist*, and *racecar driver*—are very small—with 161, 114, and 104 individuals respectively. Therefore, any change in the category *other* will also be captured by other categories. For example, there is an observed increase in the number of *tennis players* in the second half of the 1900s due to the adoption of television, but this change is already captured by the *sports* category.

### Population and technology adoption

Population data comes from two sources. At the global level we use data from the historical world population estimates of the US Census Bureau [2], which reports an aggregated dataset of world population estimates starting from the year 10,000 BC. At the city level we use a dataset on population of urban settlements from 3700 BC to AD 2000 [3]. We matched the cities present in GeoNames with the ones used in [3] based on their name, within a 30km radius. We interpolate the missing years for both population datasets using linear splines.

Data on technology adoption comes from two sources. For printing, we use the *Incunabula Short Title Catalogue* [4], a dataset comprising all books printed between 1450 and 1500. For radio and for television we use the Historical Cross-Country Technology Adoption (HCCTA) dataset [5], a dataset collected to analyze the adoption patterns of some of the major technologies introduced in the past 250 years. In particular we use the variables *Radios*, and *Televisions*. The HCCTA dataset also has historical GDP and population information. Figure **??**-B shows the fraction of countries in the HCCTA dataset with radio (solid-red) and television (dashed-blue).

## Changepoint analysis

The claim in the main text that printing coincides with a sharp increase in the per-capita number of memorable people born each year is supported by a technique used in time series analysis to detect abrupt changes in the mean of the series. The *changepoint* estimation technique [6] estimates the position and number of changepoints in a time series by assuming that the time series can be modeled by a distribution with a fixed mean. The changepoints in a time series are the points that require updating the mean of the distribution used to model the data. To find the changepoints, the technique minimizes a test statistic that depends on the number and position of the changepoints. All changepoint analyses were performed using the *changepoint* package available for R [6].

# Distance to Mainz

In this section, we will explore three different empirical specifications that allow us to establish the relation between early adoption of printing and the production of memorable scientists and artists.

First, we use the distance to Mainz as an instrumental variable for a dummy variable for whether a city adopted printing between 1450 and 1500, $D_{\text{printer}}$. As dependent variables we use the number of people born in each city between 1400 and 1550, for each occupation. Tables 1 show the result of this analysis for all occupations. First, we use the distance to Mainz as an instrumental variable for a dummy variable for whether a city adopted printing between 1450 and 1500. As dependent variables we use the number of people born in each city between 1400 and 1550, for each occupation. Tables 1 show the result of this analysis for all occupations.

Second, we use the distance to Mainz as an instrumental variable for the year of the first printed book in each city. As dependent variables we use the year of the first person born in each city, after 1400, for each occupation. Tables 2 show the results for all occupations. Here we consider only cities that adopted printing and for which we have population data.

Finally, we use the distance to Mainz as an instrumental variable for the number of printed books between 1450 and 1500. As dependent variables we use the number of people born in each city between 1450 and 1550 (right after the invention of printing in Europe) for each occupation. Tables 3 show the results of this analysis for all occupations. We consider only cities that adopted printing (i.e. $n_{\text{books}} > 0$) and for which we have population data.

| | *Dependent variable:* | | | | | | | |
|---|---|---|---|---|---|---|---|---|
| | $D_{\text{printer}}$ | | number of scientists | | number of artists | | number of political leaders | |
| | *probit* | *First Stage* | *OLS* | *IV* | *OLS* | *IV* | *OLS* | *IV* |
| | (1) | (2) | (3) | (4) | (5) | (6) | (7) | (8) |
| $\log(d_{\text{Mainz}})$ | $-0.273^{***}$ | $-0.022^{***}$ | | | | | | |
| | (0.033) | (0.003) | | | | | | |
| $D_{\text{printer}}$ | | | $0.089^{***}$ | $0.222^{***}$ | $0.202^{***}$ | $0.302^{***}$ | $0.188^{***}$ | 0.130 |
| | | | (0.004) | (0.043) | (0.007) | (0.067) | (0.009) | (0.081) |
| constant | $-0.014$ | $0.186^{***}$ | $0.002^{**}$ | $-0.003$ | $0.003^{**}$ | $-0.0002$ | $0.010^{***}$ | $0.012^{***}$ |
| | (0.212) | (0.018) | (0.001) | (0.002) | (0.001) | (0.003) | (0.002) | (0.003) |
| Observations | 5,335 | 5,335 | 5,335 | 5,335 | 5,335 | 5,335 | 5,335 | 5,335 |
| Log Likelihood | $-804.19$ | | | | | | | |
| AIC | 1,612 | | | | | | | |
| Residual SE (df = 5333) | | 0.187 | 0.061 | 0.066 | 0.100 | 0.102 | 0.124 | 0.124 |
| $R^2$ | | 0.012 | 0.070 | | 0.125 | | 0.075 | |
| Adjusted $R^2$ | | 0.012 | 0.070 | | 0.125 | | 0.075 | |
| F Statistic (df = 1; 5333) | | $66.57^{***}$ | $399.43^{***}$ | | $762.57^{***}$ | | $435.08^{***}$ | |
| *Note:* | | | | | | $^{*}$p<0.1; $^{**}$p<0.05; $^{***}$p<0.01 | | |

| | *Dependent variable:* | | | | | |
|---|---|---|---|---|---|---|
| | number of humanities | | number of religious leaders | | number of people | |
| | *OLS* | *IV* | *OLS* | *IV* | *OLS* | *IV* |
| | (9) | (10) | (11) | (12) | (13) | (14) |
| $\log(d_{\text{Mainz}})$ | | | | | | |
| $D_{\text{printer}}$ | $0.103^{***}$ | 0.058 | $0.067^{***}$ | 0.057 | $0.505^{***}$ | $0.663^{***}$ |
| | (0.006) | (0.050) | (0.005) | (0.042) | (0.014) | (0.130) |
| constant | $0.004^{***}$ | $0.005^{**}$ | $0.002^{***}$ | 0.003 | $0.024^{***}$ | $0.018^{***}$ |
| | (0.001) | (0.002) | (0.001) | (0.002) | (0.003) | (0.005) |
| Observations | 5,335 | 5,335 | 5,335 | 5,335 | 5,335 | 5,335 |
| Residual SE (df = 5333) | 0.075 | 0.076 | 0.065 | 0.065 | 0.196 | 0.199 |
| $R^2$ | 0.062 | | 0.036 | | 0.189 | |
| Adjusted $R^2$ | 0.062 | | 0.036 | | 0.189 | |
| F Statistic (df = 1; 5333) | $353.06^{***}$ | | $201.79^{***}$ | | $1,243^{***}$ | |
| *Note:* | | | | $^{*}$p<0.1; $^{**}$p<0.05; $^{***}$p<0.01 | | |

**Table 1.** Instrumental variable analysis of the effect of printing. Here, we use the distance to Mainz as an instrument for a dummy variable for adopting printing between 1450 and 1500, to estimate the effect of printing in the number of scientists, artists, political leaders, humanities, religious leaders, and all people in the dataset born between 1450 and 1550. All dependent variables are in log-scale.

| | Dependent variable: | | | | | | |
|---|---|---|---|---|---|---|---|
| | year of first printer | year of first scientist | | year of first artist | | year of first political leader | |
| | *First Stage* | *OLS* | *IV* | *OLS* | *IV* | *OLS* | *IV* |
| | (1) | (2) | (3) | (4) | (5) | (6) | (7) |
| $d_{\text{Mainz}}$ | $0.010^{***}$ | | | | | | |
| | (0.003) | | | | | | |
| year of first printer | | $4.870^{***}$ | $12.900^{**}$ | $6.552^{***}$ | $12.956^{**}$ | $3.036^{*}$ | −1.955 |
| | | (1.425) | (5.655) | (1.740) | (6.194) | (1.809) | (4.988) |
| log(*pop*) | $-3.280^{**}$ | −18.507 | −5.089 | $-37.476^{*}$ | −38.083 | $-64.311^{**}$ | $-68.651^{**}$ |
| | (1.644) | (19.035) | (24.651) | (21.950) | (24.072) | (26.015) | (27.613) |
| constant | $55.853^{***}$ | 294.220 | −69.230 | 383.895 | 215.241 | $719.812^{**}$ | $908.366^{***}$ |
| | (16.207) | (203.821) | (346.180) | (231.540) | (297.534) | (275.779) | (337.974) |
| Obs. | 82 | 73 | 73 | 70 | 70 | 77 | 77 |
| Residual SE | 10.611 | 126.760 | 152.822 | 144.475 | 158.400 | 180.853 | 189.924 |
| | (df = 79) | (df = 70) | (df = 70) | (df = 67) | (df = 67) | (df = 74) | (df = 74) |
| $R^2$ | 0.132 | 0.163 | | 0.202 | | 0.113 | |
| Adj. $R^2$ | 0.111 | 0.139 | | 0.178 | | 0.089 | |
| F Statistic | $6.032^{***}$ | $6.833^{***}$ | | $8.496^{***}$ | | $4.732^{**}$ | |
| | (df = 2; 79) | (df = 2; 70) | | (df = 2; 67) | | (df = 2; 74) | |
| *Note:* | | | | | | $^{*}$p<0.1; $^{**}$p<0.05; $^{***}$p<0.01 | |

| | Dependent variable: | | | | | | | |
|---|---|---|---|---|---|---|---|---|
| | year of first performing artist | | year of first humanities | | year of first religious leader | | year of first sports player | |
| | *OLS* | *IV* | *OLS* | *IV* | *OLS* | *IV* | *OLS* | *IV* |
| | (8) | (9) | (10) | (11) | (12) | (13) | (14) | (15) |
| $d_{\text{Mainz}}$ | | | | | | | | |
| year of first printer | $2.770^{***}$ | 3.397 | $4.327^{***}$ | 2.845 | 0.880 | 1.790 | $0.891^{***}$ | $1.273^{*}$ |
| | (0.910) | (2.675) | (1.510) | (4.076) | (2.770) | (7.350) | (0.261) | (0.678) |
| log(*pop*) | $-30.36^{**}$ | $-29.45^{**}$ | $-59.85^{***}$ | $-61.51^{***}$ | $-82.86^{**}$ | $-82.65^{**}$ | $-8.55^{**}$ | $-8.27^{**}$ |
| | (13.055) | (13.599) | (21.637) | (22.189) | (31.069) | (31.144) | (3.883) | (3.968) |
| constant | $647.2^{***}$ | $619.7^{***}$ | $705.3^{***}$ | $765.5^{***}$ | $1,073^{***}$ | $1,048^{***}$ | $556.15^{***}$ | $542.2^{***}$ |
| | (139.3) | (178.1) | (230.4) | (278.3) | (331.0) | (381.4) | (41.14) | (47.58) |
| Obs. | 71 | 71 | 77 | 77 | 48 | 48 | 74 | 74 |
| Residual SE | 87.755 | 88.061 | 151.417 | 152.399 | 185.473 | 185.695 | 26.127 | 26.519 |
| | (df = 68) | (df = 68) | (df = 74) | (df = 74) | (df = 45) | (df = 45) | (df = 71) | (df = 71) |
| $R^2$ | 0.193 | | 0.189 | | 0.139 | | 0.196 | |
| Adj. $R^2$ | 0.169 | | 0.167 | | 0.100 | | 0.173 | |
| F Statistic | $8.136^{***}$ | | $8.606^{***}$ | | $3.625^{**}$ | | $8.650^{***}$ | |
| | (df = 2; 68) | | (df = 2; 74) | | (df = 2; 45) | | (df = 2; 71) | |
| *Note:* | | | | | | | $^{*}$p<0.1; $^{**}$p<0.05; $^{***}$p<0.01 | |

**Table 2.** Instrumental variable analysis of the effect of printing on cities that adopted printing. Here, we use the distance to Mainz as an instrument for the year of the first printing press in each city, to estimate the effect of printing in the birth of scientists, artists, political leaders, performing artists, humanities, religious leaders, and sports players. The two stage least squares estimates show that cities that adopted printing earlier were the birth place of scientists and artists earlier than late adopters, but not of political leaders, performing artists, humanities, religious leaders, or sports players. Population variable corresponds to the average city population between 1400 and 1600.

| | *Dependent variable:* | | | | | | |
|---|---|---|---|---|---|---|---|
| | number of books | number of scientists | | number of artists | | number of political leaders | |
| | *First Stage* | *OLS* | *IV* | *OLS* | *IV* | *OLS* | *IV* |
| | (1) | (2) | (3) | (4) | (5) | (6) | (7) |
| $d_{\text{Mainz}}$ | $-0.948^{***}$ | | | | | | |
| | (0.314) | | | | | | |
| $\log(n_{\text{books}})$ | | $0.090^{***}$ | $0.184^{***}$ | $0.149^{***}$ | $0.235^{**}$ | $0.076^{*}$ | $-0.166$ |
| | | (0.019) | (0.068) | (0.033) | (0.105) | (0.041) | (0.152) |
| $\log(pop)$ | $1.082^{***}$ | $0.105^{*}$ | 0.029 | $0.276^{***}$ | 0.207 | $0.331^{***}$ | $0.527^{***}$ |
| | (0.303) | (0.055) | (0.081) | (0.092) | (0.125) | (0.115) | (0.181) |
| constant | $-1.735$ | $-1.191^{**}$ | $-0.730$ | $-2.893^{***}$ | $-2.471^{**}$ | $-2.959^{**}$ | $-4.147^{***}$ |
| | (3.142) | (0.545) | (0.698) | (0.916) | (1.074) | (1.148) | (1.552) |
| Observations | 81 | 81 | 81 | 81 | 81 | 81 | 81 |
| Residual SE (df = 78) | 1.908 | 0.345 | 0.394 | 0.581 | 0.607 | 0.728 | 0.877 |
| $R^2$ | 0.178 | 0.297 | | 0.346 | | 0.171 | |
| Adjusted $R^2$ | 0.157 | 0.279 | | 0.329 | | 0.150 | |
| F Statistic (df = 2; 78) | $8.458^{***}$ | $16.442^{***}$ | | $20.637^{***}$ | | $8.050^{***}$ | |
| *Note:* | | | | | | | $^{*}$p<0.1; $^{**}$p<0.05; $^{***}$p<0.01 |

| | *Dependent variable:* | | | | | |
|---|---|---|---|---|---|---|
| | number of humanities | | number of religious leaders | | number of people | |
| | *OLS* | *IV* | *OLS* | *IV* | *OLS* | *IV* |
| | (8) | (9) | (10) | (11) | (12) | (13) |
| $\log(d_{\text{Mainz}})$ | | | | | | |
| $\log(n_{\text{books}})$ | $0.084^{***}$ | 0.023 | $0.065^{***}$ | 0.017 | $0.214^{***}$ | 0.102 |
| | (0.023) | (0.075) | (0.023) | (0.073) | (0.045) | (0.144) |
| $\log(pop)$ | $0.267^{***}$ | $0.317^{***}$ | $0.189^{***}$ | $0.228^{**}$ | $0.502^{***}$ | $0.592^{***}$ |
| | (0.066) | (0.089) | (0.065) | (0.087) | (0.126) | (0.171) |
| constant | $-2.707^{***}$ | $-3.008^{***}$ | $-1.941^{***}$ | $-2.176^{***}$ | $-4.602^{***}$ | $-5.152^{***}$ |
| | (0.655) | (0.767) | (0.643) | (0.742) | (1.260) | (1.470) |
| Observations | 81 | 81 | 81 | 81 | 81 | 81 |
| Residual SE (df = 78) | 0.415 | 0.433 | 0.408 | 0.419 | 0.799 | 0.831 |
| $R^2$ | 0.347 | | 0.229 | | 0.408 | |
| Adjusted $R^2$ | 0.330 | | 0.209 | | 0.393 | |
| F Statistic (df = 2; 78) | $20.738^{***}$ | | $11.599^{***}$ | | $26.919^{***}$ | |
| *Note:* | | | | | | $^{*}$p<0.1; $^{**}$p<0.05; $^{***}$p<0.01 |

**Table 3.** Instrumental variable analysis of the effect of printing. Here, we use the distance to Mainz as an instrument for the number of books printed in each city between 1450 and 1500, restricted to cities that adopted printinge (i.e. $n_{\text{books}} > 0$). We use a two stage least squares regression to estimate the effect of printing in the number of scientists (models 2 and 3), artists (models 4 and 5), political leaders (models 6 and 7), humanities (models 8 and 9), religious leaders (models 10 and 11), and all people in the dataset (models 12 and 13), born between 1450 and 1550. All dependent variables are in log-scale.

## Number of televisions and radios

We regress the number of people born in each country in each year (between 1820 and 1998), for each category, against the number of televisions and the number of radios, controlling for GDP and population. As mentioned before, gata on GDP, population, number of radios, and number of televisions comes from the HCCTA dataset [5]. Tables 4 and 5 show the results for all occupations. The differences between the model with and without the number of radios and televisions are significant for performing arts, sports players, political leaders, and humanities (p-values of ∼ 1e-16, ∼ 1e-16, ∼ 1e-06, and ∼ 1e-03, respectively), and are not significant for scientists, artists, and religious leaders (p-values of 0.1865, 0.0308, and 0.1938, respectively). We note that the category humanities includes journalists (see Table 6).

| | *Dependent variable:* | | | | | | | | |
|---|---|---|---|---|---|---|---|---|---|
| | | number of performing artists | | | number of sports players | | | number of scientists | |
| | (1) | (2) | (3) | (4) | (5) | (6) | (7) | (8) | (9) |
| $\log(n_{\text{radios}})$ | $0.037^{***}$ | | $0.031^{***}$ | $0.027^{***}$ | | $0.029^{***}$ | 0.003 | | −0.003 |
| | (0.005) | | (0.005) | (0.004) | | (0.004) | (0.004) | | (0.004) |
| $\log(n_{\text{tvs}})$ | $0.079^{***}$ | | $0.072^{***}$ | $0.040^{***}$ | | $0.042^{***}$ | −0.001 | | −0.007 |
| | (0.009) | | (0.009) | (0.006) | | (0.006) | (0.006) | | (0.006) |
| $\log(gdp)$ | | $0.477^{***}$ | $0.312^{***}$ | | $0.091^{*}$ | −0.033 | | $0.218^{***}$ | $0.235^{***}$ |
| | | (0.069) | (0.067) | | (0.050) | (0.050) | | (0.043) | (0.044) |
| $\log(pop)$ | | $-0.283^{***}$ | $-0.158^{**}$ | | $-0.135^{**}$ | −0.041 | | −0.044 | −0.056 |
| | | (0.070) | (0.068) | | (0.053) | (0.052) | | (0.044) | (0.044) |
| constant | −0.058 | −0.268 | −0.323 | −0.107 | −0.283 | −0.385 | −0.027 | −0.514 | −0.580 |
| | (0.074) | (0.308) | (0.303) | (0.107) | (0.368) | (0.369) | (0.097) | (0.366) | (0.362) |
| Obs. | 3,011 | 3,011 | 3,011 | 3,011 | 3,011 | 3,011 | 3,011 | 3,011 | 3,011 |
| $R^2$ | 0.342 | 0.353 | 0.354 | 0.498 | 0.494 | 0.499 | 0.519 | 0.523 | 0.525 |
| Adj. $R^2$ | 0.301 | 0.313 | 0.314 | 0.467 | 0.463 | 0.468 | 0.489 | 0.494 | 0.495 |
| Residual SE | 0.331 | 0.328 | 0.328 | 0.420 | 0.421 | 0.419 | 0.386 | 0.384 | 0.384 |
| Residual SE df | 2835 | 2835 | 2833 | 2835 | 2835 | 2833 | 2835 | 2835 | 2833 |
| F Statistic | $8.408^{***}$ | $8.834^{***}$ | $8.789^{***}$ | $16.061^{***}$ | $15.836^{***}$ | $15.955^{***}$ | $17.464^{***}$ | $17.760^{***}$ | $17.692^{***}$ |
| F Statistic df | 175; 2835 | 175; 2835 | 177; 2833 | 175; 2835 | 175; 2835 | 177; 2833 | 175; 2835 | 175; 2835 | 177; 2833 |
| *Note:* | | | | | | | | | $^{*}p<0.1$; $^{**}p<0.05$; $^{***}p<0.01$ |

**Table 4.** Impact of the adoption of radio and television on the number of people born, for each category, between 1820 and 1998. All dependent variables are in log-scale, and are calculated yearly. The differences between models (2) and (3), and (5) and (6) are significant, while the difference between models (8) and (9) are not.

| | *Dependent variable:* | | | | | |
|---|---|---|---|---|---|---|
| | | number of artists | | | number of political leaders | |
| | (10) | (11) | (12) | (13) | (14) | (15) |
| $\log(n_{\text{radios}})$ | $0.007^{**}$ | | 0.003 | 0.005 | | 0.003 |
| | (0.003) | | (0.003) | (0.004) | | (0.004) |
| $\log(n_{\text{tvs}})$ | $0.014^{**}$ | | $0.010^{*}$ | $0.029^{***}$ | | $0.027^{***}$ |
| | (0.006) | | (0.005) | (0.007) | | (0.007) |
| $\log(gdp)$ | | $0.204^{***}$ | $0.184^{***}$ | | $0.114^{**}$ | 0.072 |
| | | (0.040) | (0.040) | | (0.052) | (0.052) |
| $\log(pop)$ | | $-0.096^{**}$ | $-0.081^{**}$ | | −0.050 | −0.019 |
| | | (0.040) | (0.040) | | (0.051) | (0.052) |
| constant | −0.058 | −0.268 | −0.323 | −0.107 | −0.283 | −0.385 |
| | (0.074) | (0.308) | (0.303) | (0.107) | (0.368) | (0.369) |
| Obs. | 3,011 | 3,011 | 3,011 | 3,011 | 3,011 | 3,011 |
| $R^2$ | 0.342 | 0.353 | 0.354 | 0.498 | 0.494 | 0.499 |
| Adj. $R^2$ | 0.301 | 0.313 | 0.314 | 0.467 | 0.463 | 0.468 |
| Residual SE | 0.331 | 0.328 | 0.328 | 0.420 | 0.421 | 0.419 |
| Residual SE df | 2835 | 2835 | 2833 | 2835 | 2835 | 2833 |
| F Statistic | $8.408^{***}$ | $8.834^{***}$ | $8.789^{***}$ | $16.061^{***}$ | $15.836^{***}$ | $15.955^{***}$ |
| F Statistic df | 175; 2835 | 175; 2835 | 177; 2833 | 175; 2835 | 175; 2835 | 177; 2833 |
| *Note:* | | | | $^{*}$p<0.1; $^{**}$p<0.05; $^{***}$p<0.01 | | |

| | *Dependent variable:* | | | | | |
|---|---|---|---|---|---|---|
| | | number of humanities | | | number of religious leaders | |
| | (16) | (17) | (18) | (19) | (20) | (21) |
| $\log(n_{\text{radios}})$ | 0.006 | | 0.002 | 0.003 | | 0.003 |
| | (0.004) | | (0.004) | (0.003) | | (0.003) |
| $\log(n_{\text{tvs}})$ | $0.020^{***}$ | | $0.017^{***}$ | −0.001 | | −0.001 |
| | (0.006) | | (0.006) | (0.003) | | (0.003) |
| $\log(gdp)$ | | $0.173^{***}$ | $0.146^{***}$ | | 0.010 | 0.004 |
| | | (0.045) | (0.045) | | (0.021) | (0.022) |
| $\log(pop)$ | | −0.059 | −0.039 | | −0.007 | −0.002 |
| | | (0.045) | (0.045) | | (0.021) | (0.021) |
| constant | −0.027 | −0.514 | −0.580 | $-0.051^{**}$ | −0.030 | −0.056 |
| | (0.097) | (0.366) | (0.362) | (0.020) | (0.147) | (0.148) |
| Obs. | 3,011 | 3,011 | 3,011 | 3,011 | 3,011 | 3,011 |
| $R^2$ | 0.519 | 0.523 | 0.525 | 0.181 | 0.180 | 0.181 |
| Adj. $R^2$ | 0.489 | 0.494 | 0.495 | 0.131 | 0.130 | 0.130 |
| Residual SE | 0.386 | 0.384 | 0.384 | 0.194 | 0.195 | 0.194 |
| Residual SE df | 2835 | 2835 | 2833 | 2835 | 2835 | 2833 |
| F Statistic | $17.464^{***}$ | $17.760^{***}$ | $17.692^{***}$ | $3.582^{***}$ | $3.559^{***}$ | $3.539^{***}$ |
| F Statistic df | 175; 2835 | 175; 2835 | 177; 2833 | 175; 2835 | 175; 2835 | 177; 2833 |
| *Note:* | | | | $^{*}$p<0.1; $^{**}$p<0.05; $^{***}$p<0.01 | | |

**Table 5.** Impact of the adoption of radio and television on the number of people born, for each category, between 1820 and 1998. All dependent variables are in log-scale, and are calculated yearly. The differences between models (14) and (15), and (17) and (18) are significant, while the difference between models (11) and (12), and (20) and (21) are not.

# Appendix: Table of occupations

**Table 6.** Description of Pantheon 2.0 categories and the aggregation used in our analysis.

| Occupation | Category | Number of people |
|---|---|---|
| COMIC ARTIST | ARTIST | 103 |
| GAME DESIGNER | ARTIST | 25 |
| ARTIST | ARTIST | 51 |
| PHOTOGRAPHER | ARTIST | 57 |
| DESIGNER | ARTIST | 45 |
| PAINTER | ARTIST | 917 |
| SCULPTOR | ARTIST | 105 |
| FASHION DESIGNER | ARTIST | 35 |
| ARCHITECT | ARTIST | 273 |
| COMPOSER | ARTIST | 757 |
| CRITIC | HUMANITIES | 5 |
| JOURNALIST | HUMANITIES | 83 |
| LINGUIST | HUMANITIES | 94 |
| PHILOSOPHER | HUMANITIES | 638 |
| WRITER | HUMANITIES | 3317 |
| HISTORIAN | HUMANITIES | 171 |
| GAMER | OTHER | 1 |
| SOCIAL ACTIVIST | OTHER | 306 |
| EXTREMIST | OTHER | 129 |
| YOUTUBER | OTHER | 9 |
| BADMINTON PLAYER | OTHER | 21 |
| PRODUCER | OTHER | 56 |
| MAGICIAN | OTHER | 10 |
| POKER PLAYER | OTHER | 7 |
| ATHLETE | OTHER | 1180 |
| LAWYER | OTHER | 29 |
| MOUNTAINEER | OTHER | 23 |
| MARTIAL ARTS | OTHER | 51 |
| PORNOGRAPHIC ACTOR | OTHER | 200 |
| TABLE TENNIS PLAYER | OTHER | 28 |
| PIRATE | OTHER | 18 |
| COMPANION | OTHER | 497 |
| GYMNAST | OTHER | 72 |
| EXPLORER | OTHER | 297 |
| CELEBRITY | OTHER | 137 |
| CHESS PLAYER | OTHER | 173 |
| BULLFIGHTER | OTHER | 1 |
| INSPIRATION | OTHER | 6 |
| OCCULTIST | OTHER | 5 |
| BOXER | OTHER | 92 |
| MAFIOSO | OTHER | 38 |
| HANDBALL PLAYER | OTHER | 55 |
| GOLFER | OTHER | 34 |
| CHEF | OTHER | 4 |
| MODEL | OTHER | 133 |
| ASTRONAUT | OTHER | 400 |
| SWIMMER | OTHER | 171 |
| WRESTLER | OTHER | 285 |
| TENNIS PLAYER | OTHER | 804 |
| FENCER | OTHER | 23 |
| SKIER | OTHER | 225 |
| GO PLAYER | OTHER | 2 |
| RACING DRIVER | OTHER | 666 |
| CYCLIST | OTHER | 527 |
| SKATER | OTHER | 130 |
| SNOOKER | OTHER | 22 |
| PRESENTER | OTHER | 70 |
| BUSINESSPERSON | OTHER | 326 |
| FILM DIRECTOR | PERFORMING ARTISTS | 882 |
| DANCER | PERFORMING ARTISTS | 48 |
| MUSICIAN | PERFORMING ARTISTS | 1621 |
| COMEDIAN | PERFORMING ARTISTS | 8 |
| ACTOR | PERFORMING ARTISTS | 5450 |
| SINGER | PERFORMING ARTISTS | 2175 |
| CONDUCTOR | PERFORMING ARTISTS | 92 |
| POLITICIAN | POLITICAL LEADERS | 8438 |
| PILOT | POLITICAL LEADERS | 24 |
| MILITARY PERSONNEL | POLITICAL LEADERS | 806 |
| NOBLEMAN | POLITICAL LEADERS | 287 |
| JUDGE | POLITICAL LEADERS | 27 |
| DIPLOMAT | POLITICAL LEADERS | 42 |
| PUBLIC WORKER | POLITICAL LEADERS | 13 |
| RELIGIOUS FIGURE | RELIGIOUS LEADERS | 1307 |
| PHYSICIST | SCIENTIST | 518 |
| INVENTOR | SCIENTIST | 201 |
| CHEMIST | SCIENTIST | 393 |
| BIOLOGIST | SCIENTIST | 462 |
| ECONOMIST | SCIENTIST | 218 |
| ARCHAEOLOGIST | SCIENTIST | 52 |
| GEOGRAPHER | SCIENTIST | 44 |
| ASTRONOMER | SCIENTIST | 357 |
| MATHEMATICIAN | SCIENTIST | 539 |
| PSYCHOLOGIST | SCIENTIST | 134 |
| STATISTICIAN | SCIENTIST | 9 |
| COMPUTER SCIENTIST | SCIENTIST | 141 |
| GEOLOGIST | SCIENTIST | 36 |
| ENGINEER | SCIENTIST | 226 |
| ANTHROPOLOGIST | SCIENTIST | 30 |
| PHYSICIAN | SCIENTIST | 320 |
| POLITICAL SCIENTIST | SCIENTIST | 23 |
| SOCIOLOGIST | SCIENTIST | 37 |
| VOLLEYBALL PLAYER | SPORTS PLAYERS | 47 |
| HOCKEY PLAYER | SPORTS PLAYERS | 99 |
| COACH | SPORTS PLAYERS | 115 |
| BASEBALL PLAYER | SPORTS PLAYERS | 25 |
| SOCCER PLAYER | SPORTS PLAYERS | 6482 |
| REFEREE | SPORTS PLAYERS | 65 |
| AMERICAN FOOTBALL PLAYER | SPORTS PLAYERS | 24 |
| CRICKETER | SPORTS PLAYERS | 21 |
| BASKETBALL PLAYER | SPORTS PLAYERS | 670 |
| RUGBY PLAYER | SPORTS PLAYERS | 8 |